# A wireless lingual feedback device to reduce overpressures in seated posture: A feasibility study

Olivier Chenu * ** - PhD student - corresponding author olivier.chenu@imag.fr
Nicolas Vuillerme * - PhD
Jacques Demongeot * - MD / PhD
Yohan Payan * - PhD
* Techniques for biomedical engineering and complexity management - informatics, mathematics and applications - Grenoble (TIMC-IMAG) - UMR UJF CNRS 5525 - Domaine de la Merci, 38710 La Tronche - France.
** IDS SA - 71300 Montceau les mines – France.

## Abstract

Background

Pressure sores are localized injuries to the skin and underlying tissues and are mainly resulting from overpressure. Paraplegic peoples are particularly subjects to pressure sores because of long-time seated postures and sensory deprivation at the lower limbs.

Methodology / Principal findings

Here we report outcomes of a feasibility trial involving a biofeedback system aimed at reducing buttock overpressure whilst an individual is seated. The system consists of (1) pressure sensors, (2) a laptop that couples sensors and actuator (3) a wireless Tongue Display Unit (TDU) consisting of a circuit embedded in a dental retainer with electrodes put in contact with the tongue. The principle consists in (1) detecting overpressures in people who are seated over long periods of time, (2) estimating a postural change that could reduce these overpressures and (3) communicating this change through directional information transmitted by the TDU.

Twenty-four healthy subjects voluntarily participated in this study. Twelve healthy subjects initially formed the experimental group (EG) and were seated on a chair with the wireless TDU inside their mouth. They were asked to follow TDU orders that were randomly spread throughout the session. They were evaluated during two experimental sessions during which 20 electro-stimulations were sent. Twelve other subjects were added retrospectively to form the control group (CG). These subjects participated in one session of the same experiment but without any biofeedback.

Three dependent variables were computed: (1) the ability of subjects to reach target posture (EG versus CG), (2) high pressure reductions after a biofeedback (EG versus CG) and (3) the level of these reductions relative to their initial values (EG only). Results show (1) that EG reached target postures in 90.2% of the trials, against 5,3% in the CG,

40   (2) a significant reduction in overpressures in the EG compared to the CG and (3), for the EG, that the higher the initial pressures were, the more they were decreased.

Conclusions / Significance
The findings suggest that, in this trial, subjects were able to use a tongue tactile feedback
45   system to reduce buttock overpressure while seated. Further evaluation of this system on paraplegic subjects remains to be done.

## Introduction

The tongue is currently being investigated in our laboratory as part of a biofeedback system. Its characteristics have been described by Paul Bach-y-Rita and colleagues [1] in
50   the context of sensory substitution [2]. These studies suggested that stimulus characteristics of one sensory modality (e.g., a visual stimulus) could be transformed into stimulations of another sensory modality (e.g., a matrix of vibrotactile or electrotactile stimulations on different pars of the body) [3]. For various reasons (electrical consumption of the device, potential of miniaturization and sensory characteristics of the
55   human body) Bach-y-Rita and colleagues recently converged towards electrotactile feedback devices, and particularly the electro-stimulation of the tongue [1,4]. This organ was chosen as it is the most sensitive organ in the human body, with a discrimination threshold between one and two millimeters. Bach-y-Rita and colleagues designed a Tongue Display Unit (TDU) consisting of a 2D array of miniature electrodes (12x12
60   matrix) held between the jaws and positioned in close contact with the anterior-superior surface of the tongue (Figure 1, left). A flexible cable connects the matrix to a computer-controlled external triggering device delivering the electrical signals that individually activate the electrodes.

After using a Tongue Display Unit to address pathologies such as balance dysfunction
65   [5,6], our laboratory is currently addressing the pathologies encountered by paraplegic and quadriplegic patients. The tongue is indeed important for these patients since it is usually the last organ that keeps its sensitivity and mobility after a spinal cord injury: the cranial nerves that innerve the lingual tissues often escape spinal cord injuries and are mostly unaffected by neuromuscular diseases. In the extreme case of full quadriplegics,
70   the tongue is the only way in which patients can interact with their environment (e.g. commanding the motion of their wheelchair through tongue pressure on an in-mouth embedded device [7]) or perceive tactile stimulations (via a TDU for example).

Our laboratory aims to develop a new generation of "Tongue Display Units" that would provide biofeedback information concerning the actual state of buttock pressure
75   distribution that would offer to quadriplegic patients the possibility of interacting with their environment. This new TDU would therefore offer a *passive* sensory interface coupled with *active* actuators. This paper addresses the first part of the project, namely the way overpressures artificially measured in the buttock area, can be sent and perceived by subjects through an embedded TDU.

80   Pressure sores represent a frequent but potentially preventable condition seen most often in immobilized (elderly) or neurologically impaired (paraplegic and quadriplegic) patients. Although pressure is not the only cause of pressure sore formation [8], Linder-Ganz et al. have reported a positive correlation of pressure rates and time in pressure sore formations in rats [9]. This situation may particularly occur in persons with spinal cord injuries when they are in seated posture. These patients do not get the sensory signal arising from the buttock area that allows healthy subjects to prevent pressure sores by modifying their buttock pressures. Pressure sore treatment, which can be medical or surgical, is always long, difficult and expensive. Within this context, solutions to prevent pressure sores are needed.

Different approaches have been proposed to prevent pressure sores (see Reddy et al. for a review [10]). Our group recently developed an original device with the intention of reducing overpressure in individuals with spinal cord injuries. Its underlying principle consists of (1) putting a pressure mapping system onto the wheelchair seat area that allows real-time acquisition of the pressure applied on the seat/skin interface, (2) detecting any excessive pressure concentration, (3) estimating the patient posture modification that would reduce this concentration and then (4) sending this signal to the patient. Moreau-Gaudry et al. focused on steps (1) and (4) and showed that young healthy subjects were able to correctly modify the pressure applied at the seat/skin interface, according to a random directional signal electro-stimulated on the tongue [11]. This paper focuses on the whole process, with a study that should help to evaluate the feasibility of the method by quantitatively measuring the decrease in pressure concentrations occurring after each TDU biofeedback signal.

## Materials and Methods

### Functioning of the biofeedback device

#### Detection of pressure concentrations

The Vista Medical pressure mapping system (FSA Seat 32/63 mat) was used for the acquisition of the pressure applied at the seat/skin interface. This mat is made of 1024 (32x32) pressure sensors (6.5cm² each) that sense the magnitude of pressure applied to them (Figure 2). As recommended by the manufacturer, the device was calibrated for pressures up to 200mmHg. This pressure mapping system was connected to a laptop enabling the acquisition of pressure maps (P_M) at a frequency of 5Hz.

We computed a simple algorithm for the detection of over-pressured regions: a filter removed the sensor data points whose values were below 100mmHg. Figure 3 plots the resulting high pressure distribution map (HP_M) describing the anatomical regions where the risk of pressure sores formation may be maximal. We will call this region (the ensemble of sensors above 100 mmHg) Arisk.

#### Estimation of the postural change for a reduction of overpressures

Estimating the postural change that should reduce the overpressure areas is an ill-posed inverse problem, with a number of different postures that can induce very similar pressure distributions. A forward model of the correspondence between pressure distribution and posture seems then very difficult to compute and is beyond the scope of this paper. Therefore, as a starting point, we proposed a sampling of the subject postural space to estimate, for each sample, the correspondence between sitting posture and pressure distribution at the buttocks. This sampling was defined by a 3x3 grid describing
125  the 8 subjects positions that surround the central sitting posture (Figure 4).

For each corresponding posture, the pressure distribution is measured during a short period (approximately 5 seconds). Nine prerecorded (PR) mean distribution maps are then stored: PR_FL, PR_F, PR_FR, PR_CL, PR_C, PR_CR, PR_BL, PR_B and PR_BR, corresponding to the front-left, front, front-right, center-left, center, center-right, back-
130  left, back and back-right patient mean pre-recorded maps, respectively. These maps can then be used to determine a postural change that should reduce the over-pressured areas with daily use of the device.

In a first step, the actual posture of the subject is estimated. For this, a distance between the actual pressure distribution map P_M and each pre-recorded map PR_X is computed.
135  This distance d(X)=d(P_M,PR_X) is calculated as the Manhattan distance of the Euclidian space in which each sensor defines a dimension :

$$d(A,B) = \sum_{sensors\_s} |A(s) - B(s)|$$

where A(s) (resp. B(s)) is the pressure recorded by the sensors for pressure map A (resp. B).

140  The pre-recorded map PR_X for which the distance d(X) is minimal is assumed to characterize the actual subject's posture. In a second step, the postural change that should efficiently reduce the over-pressured areas is estimated. Under the condition for which only four commands of simple postural change are given to the user (move forward, backward, left or right), postures that can be reached by the subject, starting from the
145  actual posture, are determined. For example, as illustrated in figure 4 left, if the actual posture is estimated to be back-right (BR), only the four postures CR, FR, B and BL can be reached with the postural change commands.

A weighted distance wd(Y)=wd(HP_M,PR_Y) between HP_M and the pre-recorded maps PR_Y corresponding to each reachable posture is computed :

150  $$wd(A,B) = \sum_{sensors\_s} t(A(s)) \cdot (A(s) - B(s))$$

with

t(A(s)) = max(A(s)-T, 0)

where T is a pressure truncation parameter. t(A) allows for more weight distribution to high pressures and none to pressures under T. The FSA mat reads pressures up to 200 mmHg ; according to our observations, we arbitrarily set the T value to 100 mmHg. This choice was supported by the fact that 100mmHg is inside the range of typical systolic/diastolic blood pressures. This value is easily adjustable, depending on future data.

The posture whose distance wd(Y) is the highest, is considered to be the one towards which we want to guide the subject. A high distance wd(Y) means a large decrease of the pressure inside the measured over-pressured areas.

### *Guiding the subject towards a new posture that will reduce over-pressures*

To prove the feasibility of the approach and before providing dense information onto the TDU (such as a representation of the pressure distribution), we chose to only provide four postural change commands to the user. A 6x6 TDU matrix was designed (Figure 1, center: 36 electrodes with a 0.7mm radius, embedded on a 1.5x1.5cm plastic strip, physically connected to an external electronic device). This TDU was used [11] with an activation of the matrix rows and columns corresponding to the four postural change commands, namely "move forward, backward, left or right" (figure 5). However, to provide a realistic and clinical perspective beyond the laboratory framework and to permit its use over long periods of time in a real-life environment, our device had to be ergonomically and esthetically acceptable. Of course, the ribbon TDU system did not meet these requirements. To counter this problem, we have developed the first wireless radio-controlled 6x6 TDU prototype [12]. It is inserted along with microelectronics, antenna and battery (Ion-Lithium DL2032, 220 mAh) into a bio-compatible dental retainer (figure 1, right). Each electrode can convey an electrical signal whose frequency is 50Hz and whose amplitude is adjustable between 0 and 5 volts (subject-dependent). Patterns of tongue electro-stimulations can be refreshed at about 3 Hz.

### *Feasibility study*

A feasibility study was performed in order to evaluate the effectiveness of the biofeedback device in reducing overpressure in seated posture, by quantitatively measuring the decrease in high pressure concentrations that occur after each postural change command.

The study was conducted in accordance with the Declaration of Helsinki and was approved by the local ethics committee (LPNC laboratory – CNRS). This study was registered at clinicaltrials.org under number NCT00917228.

### *Subjects*

Twenty four young healthy adults voluntarily participated in the experiment. (1) First, an experimental group (12 subjects, age = 25,8±4,2 years) was formed (2) Second, a control group was recruited retrospectively (12 subjects, age=28,9±7,4 years).

Subjects were recruited at Grenoble University. Inclusion criteria for participating were the ability to change postures in a seated position, having neither motor nor sensory disabilities, and to correctly perceive electrical stimulation on the tongue. Every volunteer was able to participate and all provided informed consent to participate.

### *Experimental procedure*

Before the experiment, TDU intensities needed to be calibrated for each subject since lingual sensibility varies between subjects (a given intensity could hurt one subject while not even be detected by another). Moreover, since the lingual receptors are non-uniformly spread on the tongue surface [13], TDU intensities needed to be calibrated separately for each pattern of stimulation (front, back, left or right). The different patterns that were displayed on the TDU are plotted on figure 5. Each subject was asked to put the signal at a high but not painful level. Then, a short TDU training procedure was carried out. Eight patterns were randomly generated and the subject had to show where he/she felt the stimulation. This was repeated until the subject correctly recognized at least seven of the eight patterns.

For the experiment, subjects were comfortably seated on a chair (in an office of our laboratory) onto which was placed the FSA pressure mapping system. This was freshly calibrated with the Vista Medical system. As described above, subjects were asked to remain still in the nine configurations that sample their postural space for several seconds, in order for the corresponding mean distribution maps to be stored.

For subjects of the experimental group, the experimentation consisted of two sessions of approximately 5 minutes separated by a pause of about 5 minutes. During each session, 20 stimulations were sent to the subject's tongue. Before each stimulation, pressure maps were recorded. After this period, the mean pressure distribution was computed, the over-pressured areas were detected, the optimal postural change was estimated and a command was sent to the subject through the wireless TDU device. If biofeedback signals were spread too regularly, subjects would possibly be able to guess when this would happen, which could falsify our measurement. The signals were, therefore, randomly spread over time. Assuming that the postural change command should simulate the sensation normally perceived when an over-pressured area appears, it was decided to send a signal that should be interpreted as a danger to avoid (i.e. the "pain signal" normally perceived by healthy subjects). For example, if the pressure distribution was concentrated at the rear part of the subject's buttock area (in which case our device may indicate a frontward postural change), then the four electrodes located at the back row of the TDU matrix would be activated. The subject would therefore try to make these stimulations disappear by moving his or her chest forward. During TDU activation, the device would keep estimating, in real-time, the actual posture of the subject. If the subject reached the target posture (or a posture towards in same direction so that he or she could not go further), the stimulation was stopped. For other postural changes (i.e. a movement that did not correspond to actual commands), the electro-stimulation was stopped after 10 seconds.

For subjects of the control group, the experimentation consisted of one session of 5 minutes under the same conditions as the experimental group, except without any biofeedback. As for the EG, a TDU signal was generated but subjects of this CG did not have the TDU device in their mouths.

### *Data Analysis*

We consider here the overpressure volume (OV) as the sum of values of sensors that belong to the region of interest Arisk.

Three dependent variables were computed:

(1) Var1 is the ability of subjects to reach the target posture (or a posture in the same direction so that he or she cannot go further) for the EG and the CG ;

(2) Var2 is the reduction of the overpressure volume (OV) after biofeedback ; this represents the global pressure decrease of risky areas in the EG and the CG ;

(3) Within the region of interest (Arisk), a more detailed analysis was conducted concerning the previous reduction Var2 measured in the EG. We proposed to separate the Arisk area into 5 classes, in which sensors were grouped by their initial pressure values (class1 grouping sensors for which initial values were between 100 and 120 mmHg, class2 120-140mmHg, class3 140-160mmHg, class4 160-180mmHg, class5 180-200mmHg). Var3 is, therefore, a multidimensional variable that gathers, for each of the five classes, the mean difference between the initial pressures and the ones after postural changes. Var3 values should allow us to estimate which sensor classes were the most affected by pressure reductions.

### *Statistical Analysis*

Analyses of variance (ANOVAs) were used for statistical comparisons. Post-hoc analyses (pre-planned comparisons) were used whenever necessary. The level of significance was set as 0.05.

## Results

Figure 6 illustrates an example of the mean pressure distribution recorded before (Figure 6a) and after (Figure 6g) the command for a postural change. In this case, after filtering the high pressures (Figure 6c), the area Arisk located around the subject's right ischium appears to be over-pressured. The corresponding command sent to the TDU is the activation of the right column of the matrix (figure 6f), which, in turn, stimulates the subject to move his or her chest from right to left. A significant reduction of the pressure inside Arisk is then observed (figure 6i), which was the objective of the pressure sores prevention device. Figure 6e plots the estimation of the actual subject posture before TDU activation, as well as the target posture. Figure 6f shows the TDU electrodes activation while the resulting change in the subject's posture is illustrated in figure 6h.

In order to evaluate the efficiency of the pressure sores prevention device more quantitatively, three dependent variables were computed.

### Var1: Postural behavior

Following what was already measured by Moreau-Gaudry and colleagues with a simplified pressure mapping system and a wire TDU matrix [11], the postural change provided by the subject after TDU activation was first assessed. The underlying idea consisted of checking whether subjects actually perceived the TDU activation and understood the command by responding with an appropriate postural change. For each TDU command, subject postures were therefore estimated before and after activation enabling a comparison between the directional command and the center of pressure movement. As illustrated in figure 7a, only 5.3% of the CG trials reached the correct posture, while respectively 90.4% and 90% of the EG1 and EG2 trials reached it. Focusing on the global EG trials, we observe that 90.2% of the trials reached the correct posture, 5.4% characterized movements that were in the appropriate direction but stopped before reaching the postural target (for example, with a back left starting posture, a back right target posture and a back center final position), 2.7% corresponded to no postural change (subjects probably did not perceive the stimulation), while 1.7% described a movement in the wrong direction (false interpretation of the electro-stimulation command).

### Var2: Overpressure Variations

Data obtained for the OV measured before and after the TDU activation were submitted to a 2 Groups (Control versus Experimental) $\times$ 2 Temporal frames (Before versus After TDU activation) ANOVA with repeated measures on the last factor. The ANOVA showed a main effect of Temporal frame ($F(1,22)=15.98$, $P<0.001$) and a significant interaction Group $\times$ Temporal frame ($F(1,22)=12.54$, $P<0.01$).

As illustrated in Figure 7b, the decomposition of this interaction into its primary main effects showed that: (1) for the Control group, no significant difference in OV was observed before and after TDU activation ($P>0.05$), whereas (2) for the Experimental group, TDU activation yielded a decreased OV ($P<0.0001$).

### Var2: Effect of practice

Data obtained for the OV measured before and after the TDU activation in the Experimental group were submitted to a 2 Experimental sessions (First versus Second) $\times$ 2 Temporal frames (Before versus After TDU activation) ANOVA with repeated measures on both factors. The ANOVA showed a main effect of Temporal frame ($F(1,11)=13.99$, $P<0.01$), yielding a decreased OV After relative to Before the TDU activation. As illustrated in Figure 7b, the absence of interaction Sessions $\times$ 2 Temporal frame ($F(1,11)=0.08$, $P>0.05$) further showed that the TDU activation yielded a similar effect on the OV in both the First and the Second experimental sessions.

### Var3: Overpressure Variations depending on initial pressures

Data obtained for the absolute reduction of overpressure volume after TDU activation for the five predetermined ranges of original pressure volume were submitted to another Student's t-test analysis (5 Initial pressure volumes (100/120 vs. 120/140 vs. 140/160 vs. 160/180 vs. 180/200 mmHg)). Results showed a main effect of Initial pressure volume (F (4,40) = 17.11, P < 0.001, Figure 7 right), yielding larger reductions of the pressure volume (1) in the 120/140 than 100/120 mmHg range (P < 0.05), (2) in the 140/160 than 120/140 mmHg range (P < 0.05), and (3) (1) in the 160/180 than 140/160 mmHg range (P < 0.05). In other words, the higher the initial pressure volumes were (i.e., before the TDU activation), the more they were decreased following the TDU activation. These results are shown in figure 7c.

## Discussion

The goal of this work was to assess if a biofeedback device could help subjects to reduce overpressures in a seated posture. A closed-loop biofeedback system was developed with the aim of changing the postural behavior of subjects. It consisted of a pressure mapping system and a wireless lingual electro-tactile device, both linked together by a laptop. The principle was to detect regions of overpressure and to send an electro-stimulation on the tongue via the TDU in order to tell the subject to move his or her chest towards a posture whereby the overpressure would be reduced. The experiment by Moreau-Gaudry et al. [11] concluded that young healthy subjects could correctly interpret an electrical lingual biofeedback to adapt their postural behavior.

This paper focused on three new points: (1) closing the biofeedback loop by considering previous data before sending the directional order, (2) using a wireless TDU which is more ergonomically and esthetically acceptable, and (3) analyzing reductions of high pressures after the TDU activation.

The results of our experiment confirmed those obtained by Moreau-Gaudry et al. [11], in that subjects were able to correctly interpret the directional signal provided by the wireless TDU. Interestingly, our results further showed that these postural changes significantly reduced the pressure in certain regions estimated to be risky for the formation of pressure sores. These regions of high pressure are localized around ischia, which are one of the more frequent areas where pressure sores may appear [14]. Inside the region of interest, the higher the pressure were before electro-stimulation, the more they decreased after electro-stimulation This result is encouraging since high pressure regions are far away from a safe pressure level, but their changes during the biofeedback make them reach a safe level more quickly. However, if the objective of pressure decrease inside Arisk is reached, it is important to note that new high pressures can appear in other regions of the buttock as a direct consequence of the subject's postural change. These high pressures (if they persist) should be corrected by the device during the next tongue electro-stimulation.

One could presume that these postural behaviors were natural and not due to the feedback device. Indeed, subjects were not disabled and their natural feeling could have led them to modify their posture. However, our results show that subjects of the control group didn't move in 87,7% of the trials. They happened to move and reach the correct posture in 5,3%, but did move in the wrong direction in 7% of the trials (indicating that the direction they move towards is probably random). Moreover, the overpressure decrease is significantly higher in the experimental group than in the control group.

Furthermore, as illustrated in figure 5, medio-lateral displacements seem more effective than antero-posterior ones : 84,79% of optimal posture modifications decided by our algorithm were medio-lateral. This observation could be relevant for conducting future studies.

Of course, the relatively simple signals corresponding to each TDU pattern could easily be conveyed by other means that could be judged less intrusive. If we think that visual or acoustic biofeedback should be avoided as they may disturb the subject in his or her daily life interaction with others, we could indeed propose simpler tactile stimulation patterns on other parts of the body which could be explored in the future. However, this feasibility study represents only the first part of a much larger project aiming at sending rich signals to the TDU. It is known, thanks to Bach-y-Rita experiments on vision [3], that subjects can interpret complex information via the TDU. The effectiveness of displaying a 6x6 filtered pressure map has to be studied.

The present findings show evidence of the ability of young healthy adults to use a tongue biofeedback to reduce overpressures of the buttock in a seated posture. In that sense, they confirm previous studies of closed-loop biofeedback systems using the TDU in many applications: vision [15], image guided surgery [16], proprioception [17] and balance control [5,18]. We believe that the ergonomic qualities of this device are numerous: high sensitivity and discrimination properties of the tongue, low power consumption, usage of a less recruited sensory channel, possibility of displaying more complicated signals and the perspective of converging towards a sensory-motor device.

As a next step, our laboratory is carrying out a clinical study involving paraplegic participants in order to validate the methodology for patients lacking any sensory feedback in the buttock area and with reduced mobility.

## Acknowledgments
The authors would like to thank the volunteers for giving a few of their time and their instructive remarks; Coronis-Systems and Guglielmi Technologies for helping in the TDU wireless design ; VistaMedical for providing the pressure mapping system ; IDS SA Company and fondation Garches for all their supports ; Nikolaï H for english correction ; Stéphanie B for various contribution.

**Figure Legends**

Figure 1. Different versions of the TDU. Left: the generic version of 12x12 electrodes; center: A miniaturized version of 6x6 electrodes; right: our wireless version (6x6 electrodes)

Figure 2. Pressure map. the VistaMedical Ltd. pressure mapping system connected to a laptop.

Figure 3. Sample of a posture and pressure map. Left: Example of a seated posture (front-right position of the chest); center: measured pressure distribution; right: filtered pressure distribution (only pressures above 100 mmHg).

Figure 4. Relation between postures and pressure distributions. Sample of pressure distributions (right) corresponding to nine seated postures (left).
 Accessible postures: From actual posture (BR for instance), only four pre-recorded postures (CR, FR, B and BL) are reachable by a simple unidirectional movement (forward or left in this case).

Figure 5. Patterns of electrostimulations. Electrodes in the front, back, left or right part of the tongue are activated when a displacement of the chest respectively in the back, front, right and left direction is suitable.

Figure 6. General principle of biofeedback and behavior of the subject. (a) Mean pressure distribution after a random timing; (b) Color values; (c) Pressures into the region of interest (pressures above 100 mmHg); (d) Pressure map at the beginning of the stimulation; (e) Estimated posture in red (corresponding to map (d)) and target posture estimated to be safe in green; (f) Directional order sent on the TDU; (g) Pressure map measured at the end of stimulation; (h) Position at the end of the stimulation, fitting the target one (in red); (i) Pressure in the region of interest at the end of the stimulation.

Figure 7. Results. (a) Global results: Postural behaviors for all trials in percentage: in white, correct behaviors; in light gray, displacement in the correct direction but not far enough ; in dark gray, no movement ; in black, movements in the wrong direction.
(b) Mean and standard deviation of the overpressured volume (OV) of the region of interest Arisk before and after electrostimulation observed (*: $p<0.05$); (c) Mean and standard deviation of pressure reduction for each predetermined range of high pressures (*: $p<0.05$).

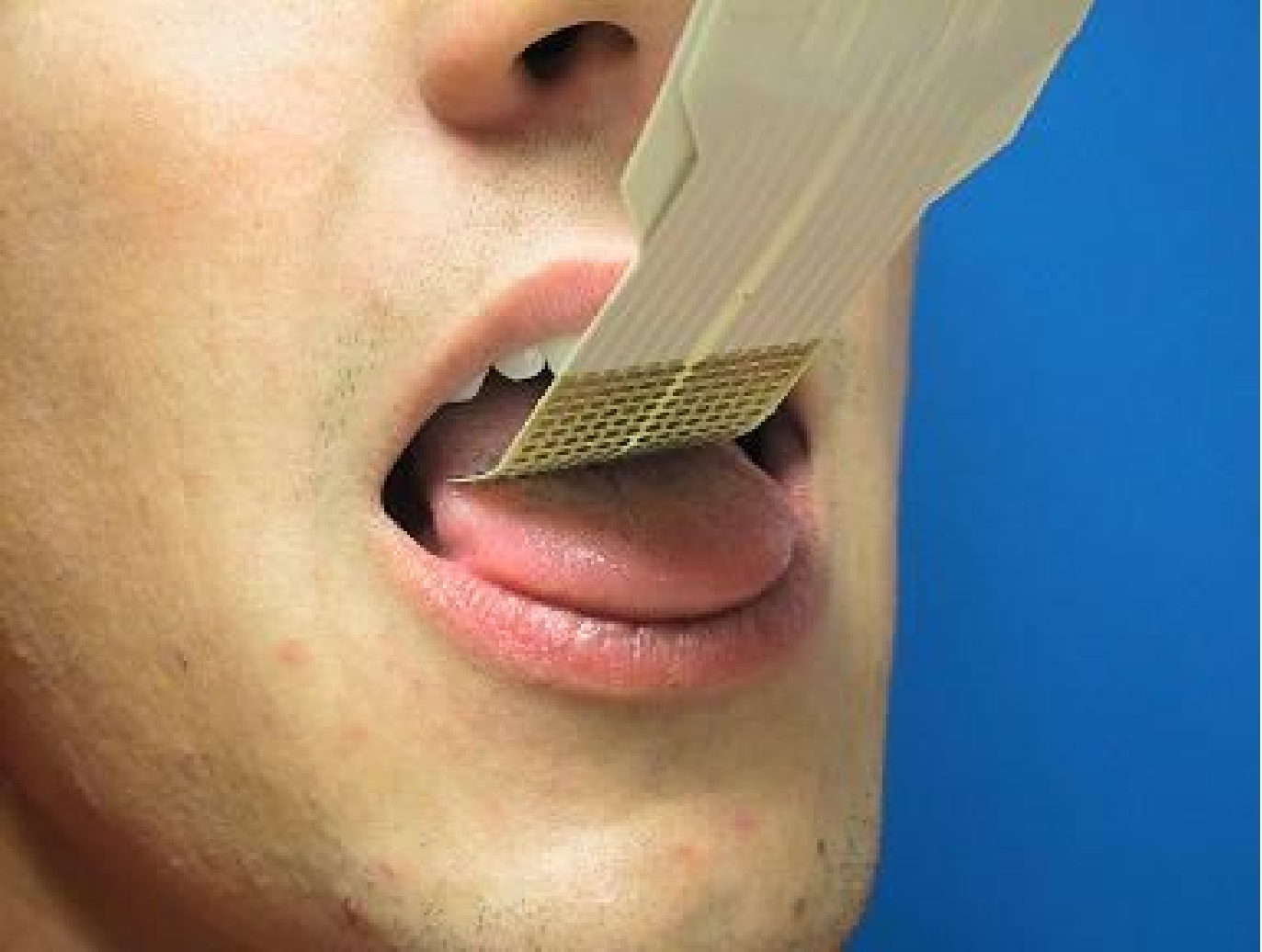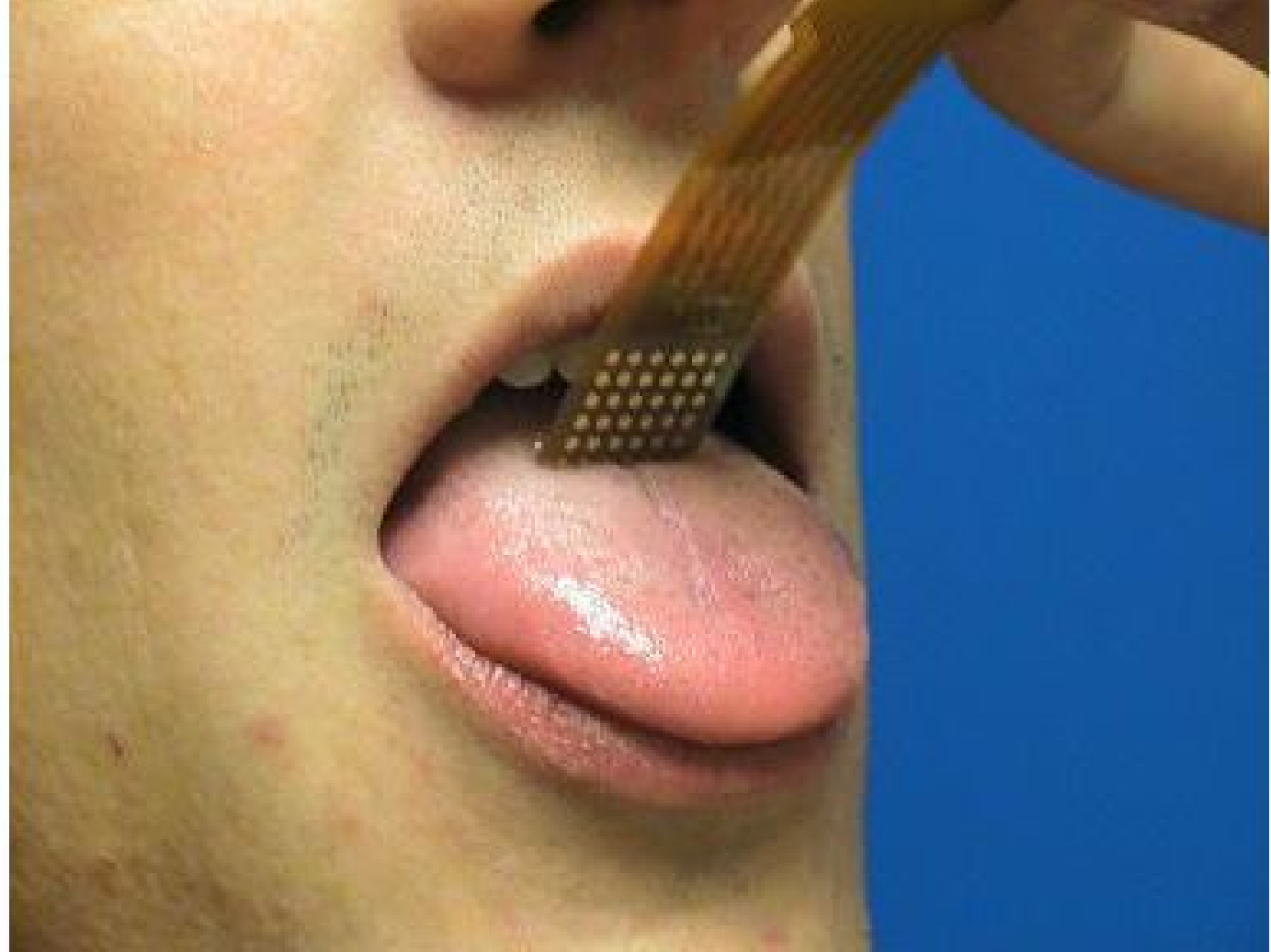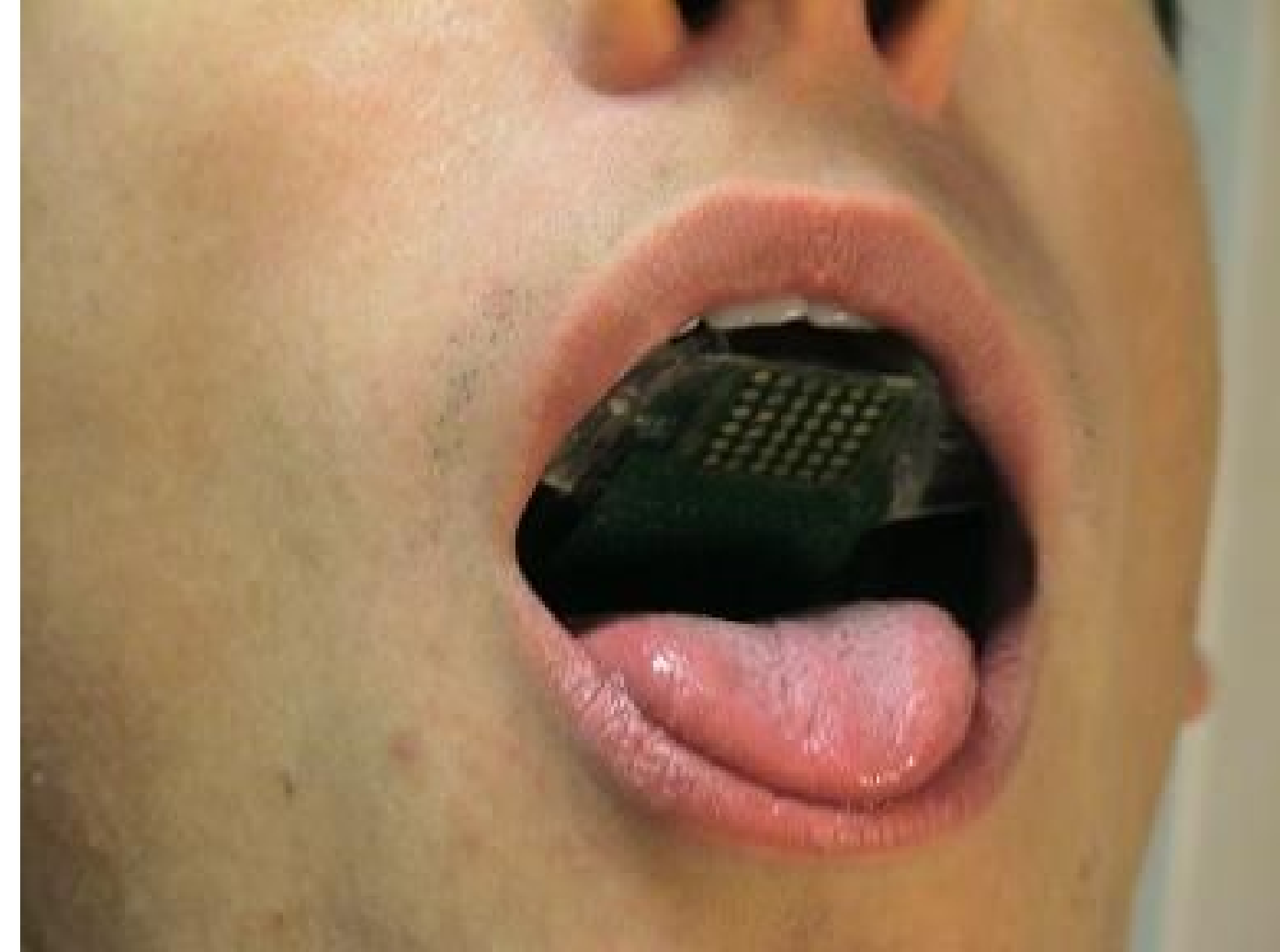

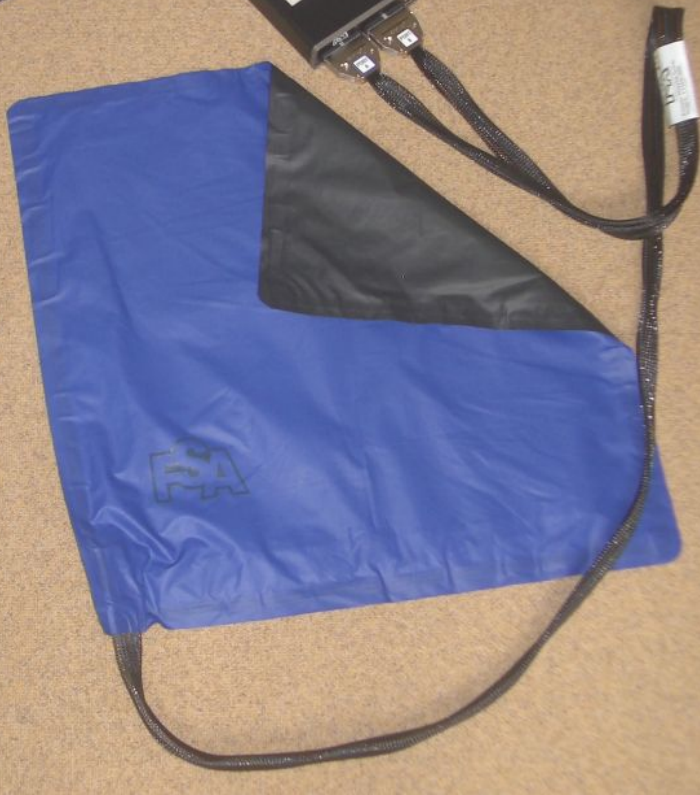

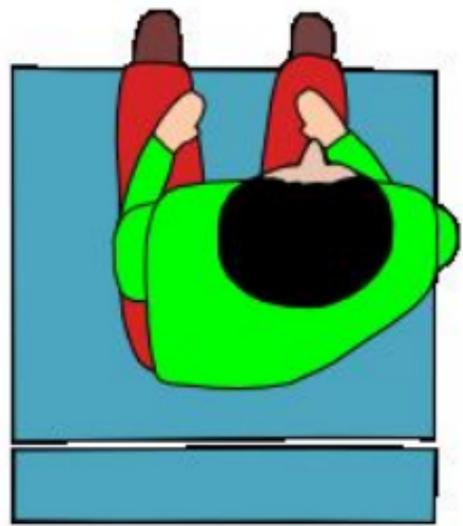 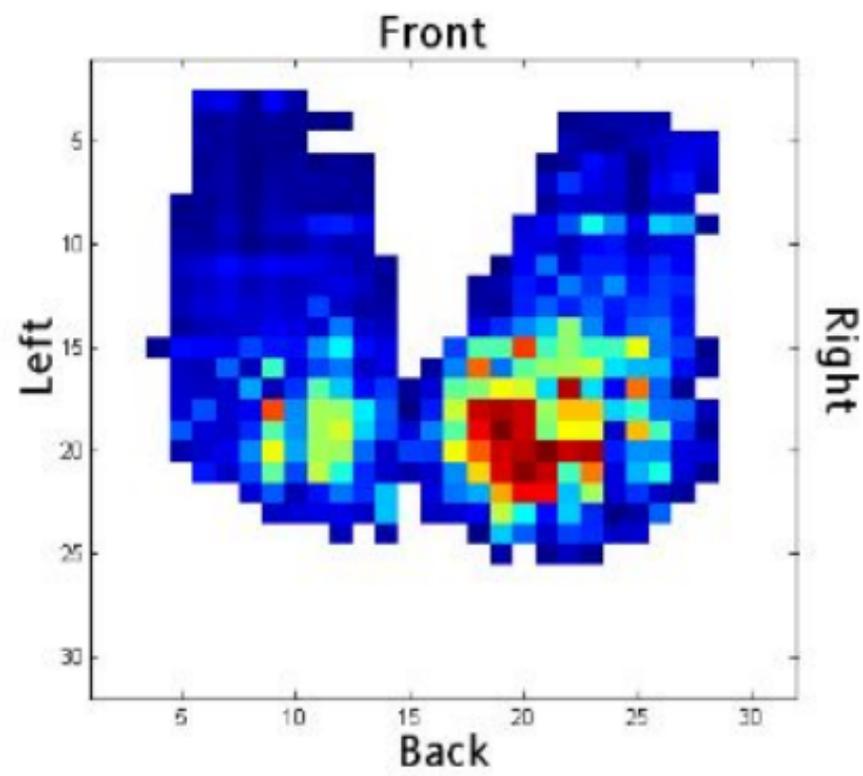 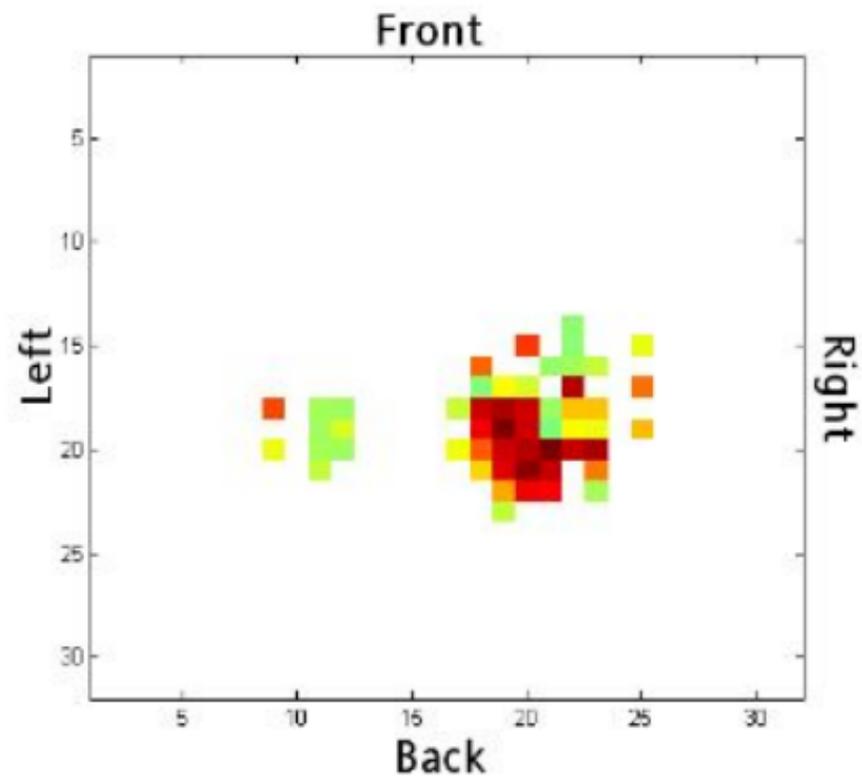

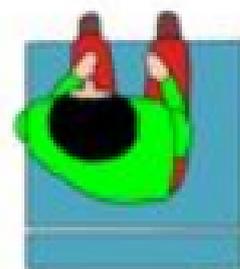 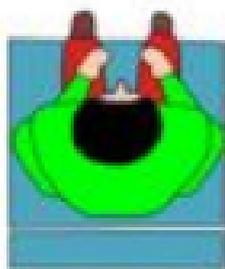 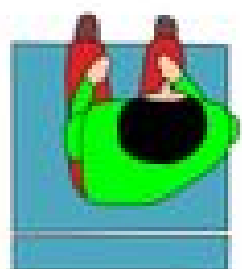 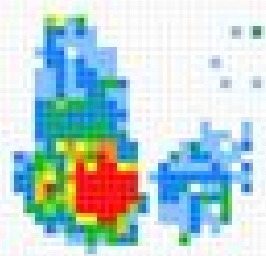 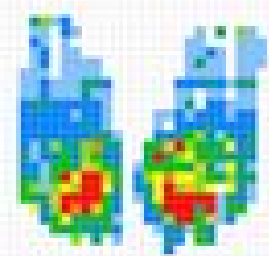 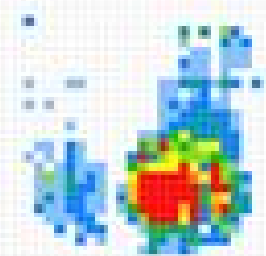
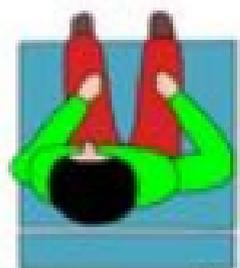 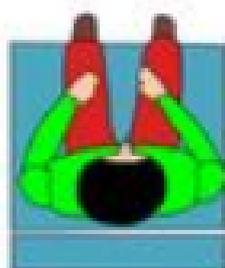 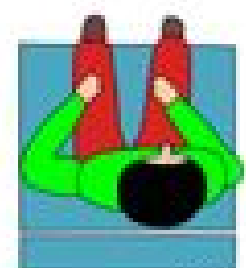 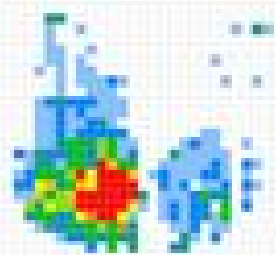 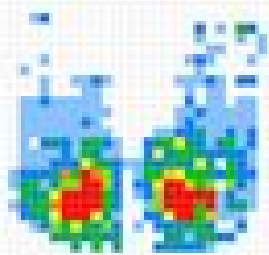 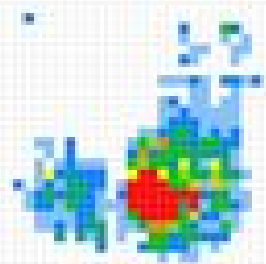
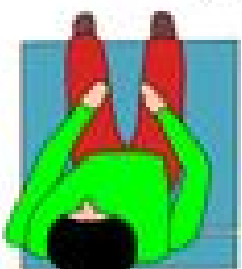 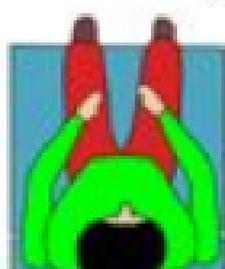 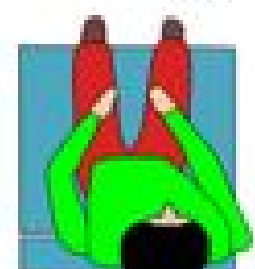 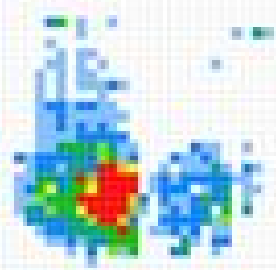 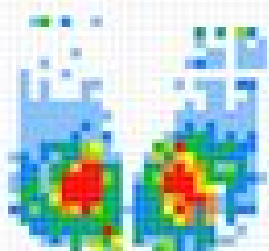 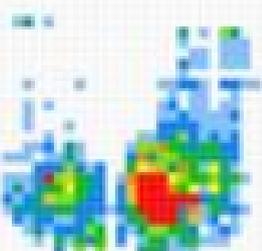

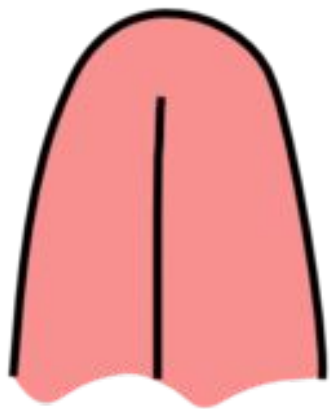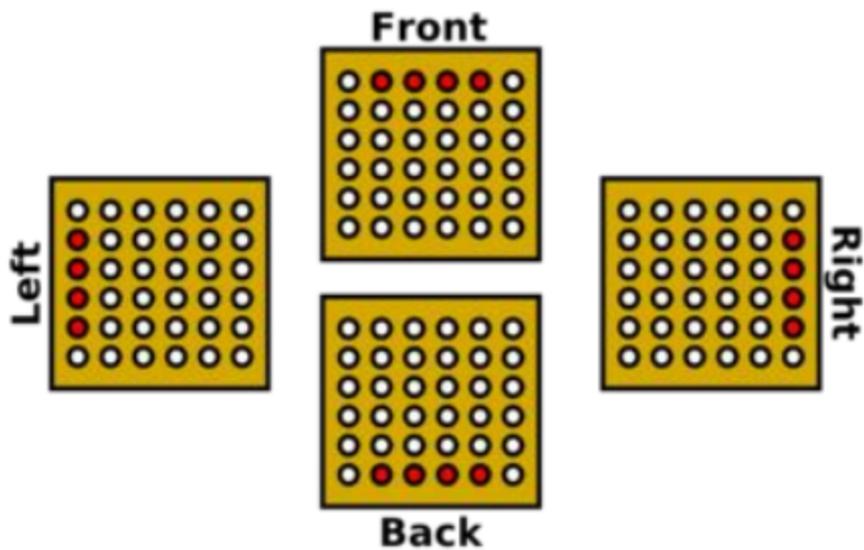

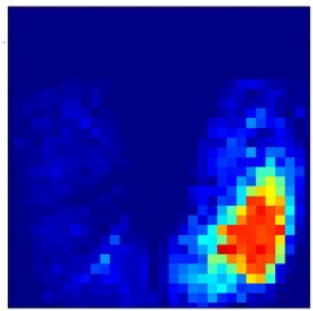 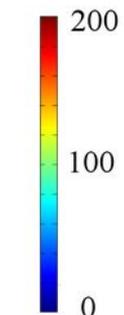 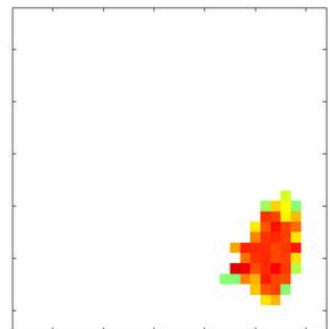

(a) (b) mmHg (c)

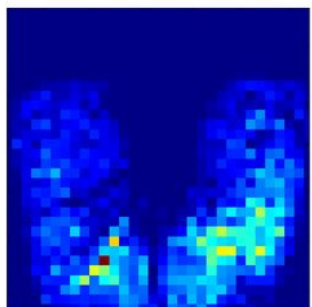 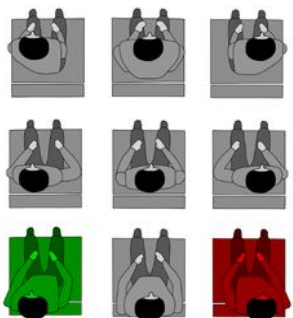 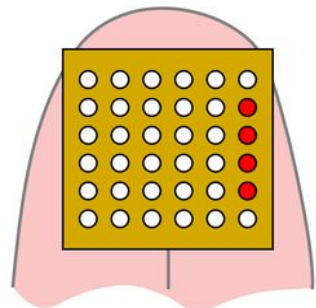

(d) (e) (f)

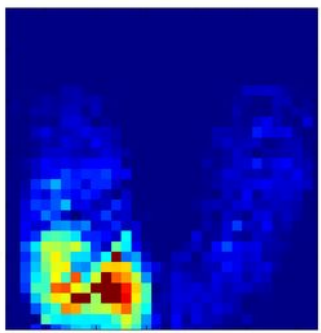 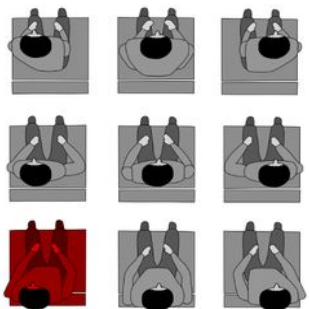 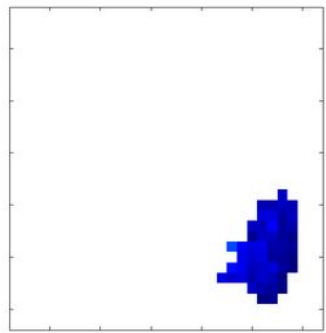

(g) (h) (i)

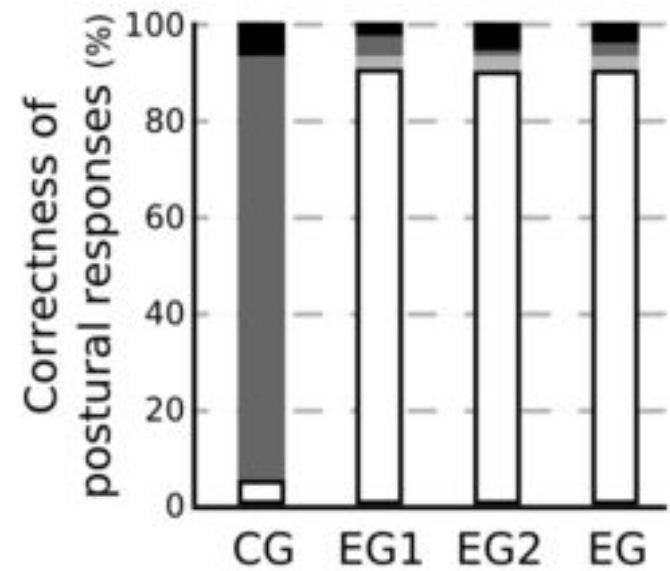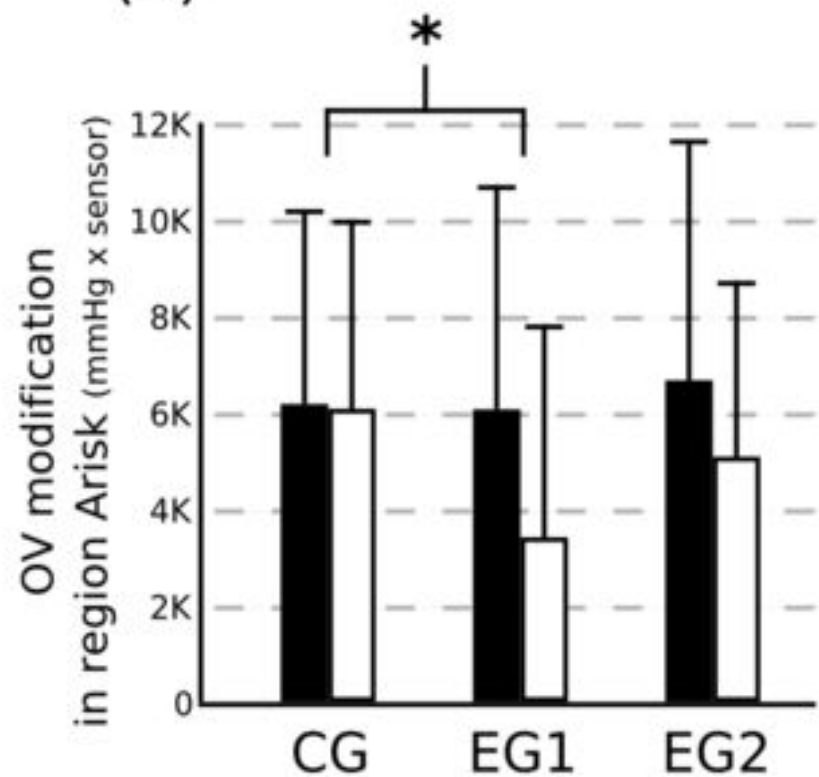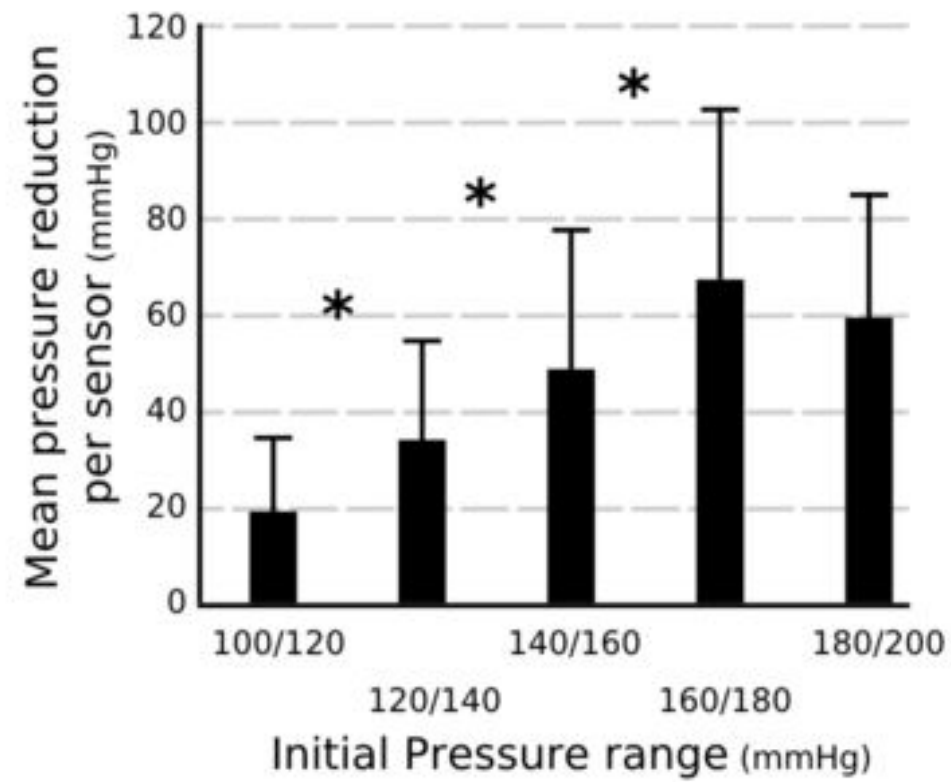